\documentclass[a4paper,11pt]{article}
\pdfoutput=1 

\usepackage{jcappub} 

\usepackage[T1]{fontenc} 

\usepackage{Extrapackages}
\usepackage{xcolor}
\usepackage{cancel}

\usepackage{arydshln}

\usepackage{etoolbox}
\makeatletter
\patchcmd{\ttlh@hang}{\parindent\z@}{\parindent\z@\leavevmode}{}{}
\patchcmd{\ttlh@hang}{\noindent}{}{}{}
\makeatother

\title{\boldmath{Local, algebraic simplifications of Gaussian random fields}}

\author{Theodor Bjorkmo}
\author{and M. C. David Marsh}

\affiliation{Department of Applied Mathematics and Theoretical Physics, University of Cambridge\\Wilberforce Road, Cambridge, UK}

\emailAdd{t.bjorkmo@damtp.cam.ac.uk}
\emailAdd{m.c.d.marsh@damtp.cam.ac.uk}

\abstract{Many applications of Gaussian random fields and Gaussian random processes are limited by the computational complexity of evaluating the probability density function, which involves inverting the relevant covariance matrix. In this work, we show how that problem can be completely circumvented for the local Taylor coefficients of a Gaussian random field with a Gaussian (or `square exponential') covariance function. Our results hold for any dimension of the field and to any order in the Taylor expansion. We present two applications. First, we show that this method can be used to explicitly generate non-trivial potential energy landscapes with many fields. This application is particularly useful when one is concerned with the field locally around special points (e.g.~maxima or minima), as we exemplify by the problem of  cosmic `manyfield' inflation in the early universe. Second, we show that this method has applications in machine learning, and greatly simplifies the regression problem of determining the hyperparameters of the covariance function given a training data set consisting of local Taylor coefficients at  single point. 
An accompanying Mathematica notebook is available at \href{https://doi.org/10.17863/CAM.22859}{https://doi.org/10.17863/CAM.22859}.

}

\begin{document}

\maketitle

\flushbottom

\newcommand{\Mpl}{M_{\mathrm{Pl}}}

\newcommand{\Nf}{{N_{\mathrm{f}}}}
\newcommand{\Lh}{\Lambda_{\mathrm{h}}}
\newcommand{\Lv}{\Lambda_{\mathrm{v}}}
\newcommand{\epsilonV}{\epsilon_{\mathrm{V}}}
\newcommand{\etaV}{\eta_{\mathrm{V}}}
\newcommand{\epsilonH}{\epsilon_{\mathrm{H}}}
\newcommand{\etaH}{\eta_{\mathrm{H}}}
\newcommand{\epsiloni}{\epsilon_{\mathrm{i}}}
\newcommand{\etai}{\eta_{\mathrm{i}}}

\newcommand{\grf}{f}
\newcommand{\pos}{\vv x}

\newcommand{\vv}[1]{\mathbf{#1}}
\newcommand{\vh}[1]{\mathbf{\hat{#1}}}
\newcommand{\nb}{\mathbf{\nabla}}
\newcommand{\dv}{\mathbf{\nabla}\cdot}
\newcommand{\cl}{\mathbf{\nabla}\times}
\newcommand{\bs}[1]{\boldsymbol{#1}}
\newcommand{\vvv}{\mathbf{v}}
\newcommand{\vvu}{\mathbf{u}}
\newcommand{\vvx}{\mathbf{x}}
\newcommand{\vvk}{\mathbf{k}} 
\newcommand{\vvp}{\mathbf{p}} 
\newcommand{\vvr}{\mathbf{r}}
\newcommand{\vvR}{\mathbf{R}} 
\newcommand{\vvG}{\mathbf{G}} 
\newcommand{\mc}[1]{\mathcal{#1}} 
\newcommand{\vvrd}{\dot{\mathbf{r}}}
\newcommand{\vvrdd}{\ddot{\mathbf{r}}}
\newcommand{\rrd}{\dot{r}}
\newcommand{\rrdd}{\ddot{r}}
\newcommand{\thetad}{\dot{\theta}}
\newcommand{\thetadd}{\ddot{\theta}}
\newcommand{\phid}{\dot{\phi}}
\newcommand{\phidd}{\ddot{\phi}}

\newcommand{\vvxs}{\mathbf{x}_\star}

\newcommand\be{\begin{equation}}
\newcommand\ee{\end{equation}}
\newcommand{\bea}{\begin{eqnarray}}
\newcommand{\eea}{\end{eqnarray}}

\newcommand{\eq}[1]{\begin{equation}{#1}\end{equation}}
\newcommand\nt{\addtocounter{equation}{1}\tag{\theequation}}
\newcommand{\mui}{^{\mu}}
\newcommand{\mli}{_{\mu}}
\newcommand{\nui}{^{\nu}}
\newcommand{\nli}{_{\nu}}
\newcommand{\pd}[2]{\frac{\partial {#1}}{\partial {#2}}}
\newcommand{\td}[2]{\frac{d{#1}}{d{#2}}}
\newcommand{\spd}[2]{\frac{\partial^2 {#1}}{\partial {#2}^2}}
\newcommand{\std}[2]{\frac{d^2 {#1}}{d{#2}^2}}

\section{Introduction}
Gaussian random fields (GRFs) are a simple class of random functions with many important applications in mathematics, computer science, and the natural sciences. A real-valued and stationary GRF, $f({\bf x})$: $ \mathbb{R}^d \to \mathbb{R}$, is completely described by its mean value  $\bar f$ and the covariance function $C(\vvx_1-\vvx_2)$:
\begin{equation}
\langle (\grf(\vvx_1)-\bar \grf)(\grf(\vvx_2)-\bar \grf)\rangle=C(\vvx_1-\vvx_2) \, .
\label{eq:GRF1}
\end{equation}

Two important problems for the practical use of GRFs are finding methods for: {\it i)} generating explicit realisations of the GRF given a specification of $\bar f$ and $C(\vvx_1-\vvx_2)$; {\it ii)} constraining  $\bar f$ and $C(\vvx_1-\vvx_2)$, given their (hyper-)parametrisation  
and some initial data of a realisation $f$.

The first problem has a a number of applications in physics,  where, for example, explicit realisations of $f$ can provide random initial conditions to physically interesting partial differential equations (cf.~e.g.~\cite{Bardeen:1985tr}). 
Naively, one may approach this problem by  sampling  $f$ for a set of points $\vvx_1,\ldots,\vvx_M$ in $\mathbb{R}^d$,
for which the  sample set $\mathcal{S} = \{f(\vvx_1),\ldots,f(\vvx_M)\}$   consists of Gaussian random variables with covariances determined from equation \eqref{eq:GRF1}. However, since the probability distribution for ${\cal S}$ depends on the inverse of the corresponding covariance matrix, this method becomes computationally very challenging when  $M \gg 1$. A common, alternative approach is instead to generate $f$ through its Fourier coefficients, which are statistically independent. This method can be efficient even for a wide range of scales, but it is not well-suited for all problems. In particular, it does not easily generalise to the conditional problem of generating $f$ given that $f(\vvxs)$ has some special properties (e.g.~is a maximum, minimum or a saddle-point).

The second problem is an important part  of model selection for Gaussian processes,\footnote{We follow the convention of e.g.~\cite{MachineLearning} and use GRFs and Gaussian processes synonymously (in particular, both may refer to any $d\geq1$).}  which has important applications  in machine learning. 
Given the training data set ${\cal S}$, one would like to make predictions for the function $f(\vvx)$ beyond the sampled points. 
In the Bayesian approach, this problem is often usefully approximated by maximising the marginal likelihood with respect to the hyperparameters of $\bar f$ and $C(\vvx_1-\vvx_2)$. 
The bottleneck of this method arises from the need to invert the corresponding covariance matrix, which becomes intractable when the training data set is very large. Moreover, pristine  training data is sometimes only easily obtained in a limited region in $\mathbb{R}^d$.

In this paper, we present a novel method for addressing these problems for the special but frequently considered case in which the GRF is stationary, isotropic and with a Gaussian (or  squared exponential) covariance function:
\begin{equation}
C(\vvx_1-\vvx_2)=h^2\, {\rm exp}\left(-\tfrac{(\vvx_1-\vvx_2)^2}{2\ell^2}\right)\, .
\label{eq:Gcovar}
\end{equation}
The hyperparameters of this covariance function are $h$ and $\ell$; we take $\bar f=0$ but our results generalise straightforwardly to $\bar f\neq 0$. Our method is {\it local}:
for problem {\it i)} we construct the explicit realisation of  $f$ through its Taylor expansion up to the order $n_{\rm max}$ around a single  point  $\vvxs\in\mathbb{R}^d$. By specifying some of the coefficients by hand, this method can be used to conditionally generate $f$ given that it has e.g.~a maximum, minimum or saddle point at $\vvxs$, but is otherwise random.  
For problem {\it ii)} we consider initial data of $f$ to be of the form of such Taylor coefficients. Both problems now depend on the inverse covariance matrix of the Taylor coefficients at $\vvxs$, which becomes large when $d\gg1$ or $n_{\rm max}\gg1$.

\begin{figure}
    \centering
    \includegraphics[width=0.55\textwidth]{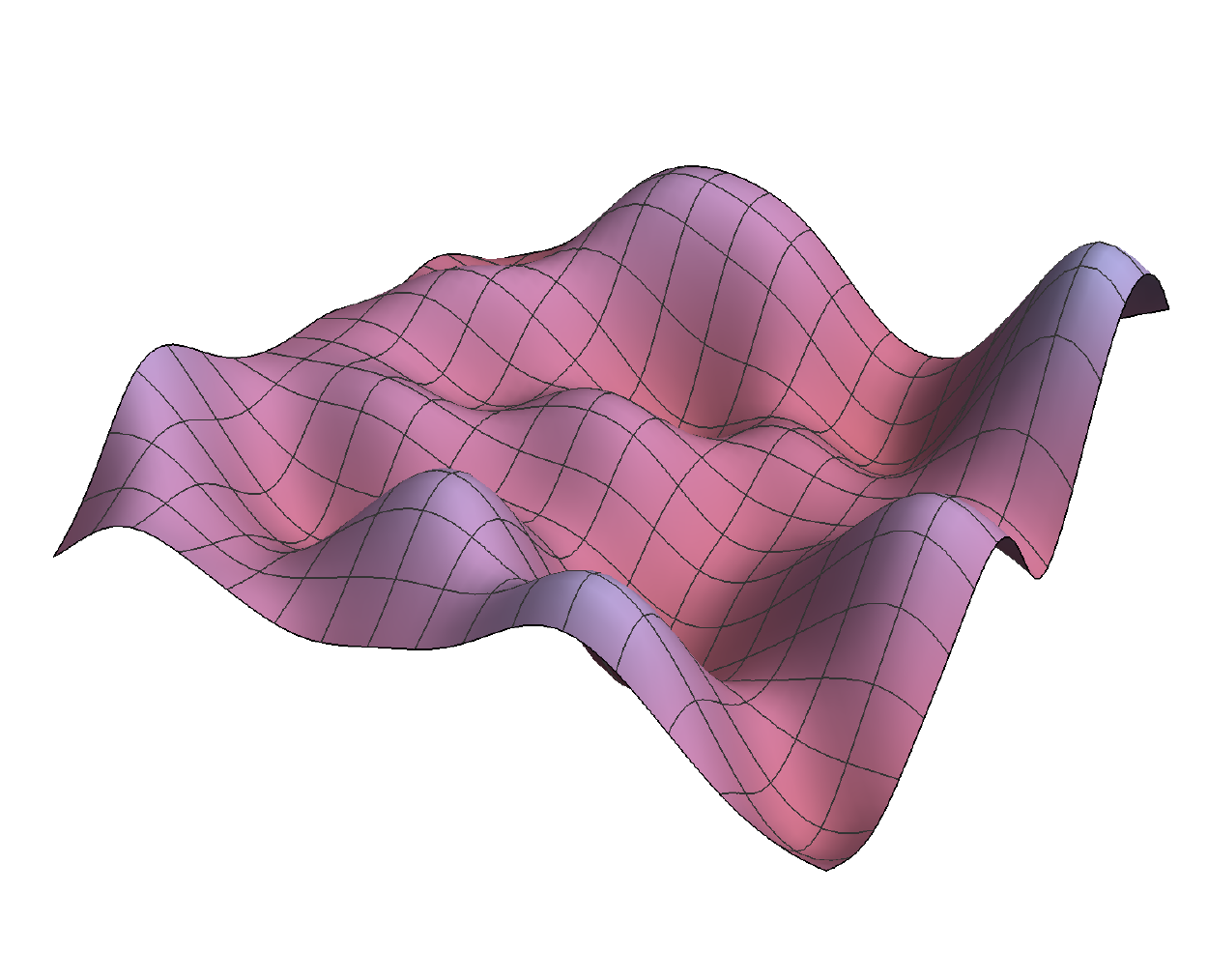}
    \caption{An example of a GRF with $d=2$ and $n_{\rm max} =175$. Here $\vvx \in [-4l,~4l]^2$.}
    \label{fig:2dpotential}
\end{figure}

The key simplification of our method is that the covariance matrix becomes exactly diagonal through a sequential, order-by-order use of the marginal and conditional probability distributions for the Taylor coefficients. This dispenses with the need to numerically invert the covariance matrix, which makes our method efficient even for large $d$ or $n_{\rm max}$.

We expect this method to be particularly useful 
when one is interested in  generating explicit GRFs  within a moderately small region in $\mathbb{R}^d$ (say, extending at most a few units of $\ell$ from $\vvxs$), 
and when one wants to constrain $h$ and $\ell$ from essentially noise-free initial data  obtained close to $\vvxs$.

 In reference
 \cite{Bjorkmo:2017nzd}, we showed that this method enables the first explicit  studies of cosmic inflation in models with a large number of fields 
interacting through a potential energy modelled by a GRF (for related work, see \cite{Tegmark:2004qd, Frazer:2011tg, Frazer:2011br, Bachlechner:2014rqa, Masoumi:2017xbe, Blanco-Pillado:2017nin}). 
For  $n_{\rm max} =5$ (motivated physically in \cite{Bjorkmo:2017nzd}),  a model with $d=100$ fields involve 96,560,546 interaction terms, and explicitly generating the potential naively involves diagonalising a covariance matrix with ${\cal O}(10^{15})$ independent entries. Our method makes this problem tractable by  allowing us to trivialise the inversion, 
condition $f$ on that the potential is suitable for inflation,
and only generate statistically independent Gaussian random numbers. Similarly, our method is efficient for moderate $d$ but $n_{\rm max} \gg1$, as illustrated by an example in Figure \ref{fig:2dpotential}. 

Future extensions of this method may, for example,  simplify the application of Gaussian process techniques to certain non-linear dynamical systems, where training data is often most easily obtained locally around equilibrium points.

This paper is organised as follows: in \S\ref{sec:method} we first illustrate the relevant aspects of our method with a simple example, and then prove its validity for any $d$ and $n_{\rm max}$. In \S\ref{sec:apps} we briefly discuss how this method can be used to address the two problems mentioned in this introduction. We conclude in \S\ref{sec:concl}.

\section{A new method for generating realisations of a GRF}
\label{sec:method}

In this section, we first review some relevant properties of GRFs and establish our notation. We then exemplify our method by considering $d=2$ and $n_{\rm max} =4$, before turning to the general, recursive proof of the method. 

A standard result, which will be our starting point, is that the covariances for the derivatives of $f$ are given by the derivatives of the covariance function,
\begin{equation}
\left\langle \frac{\partial^{n_1}\grf(\vvx_1)}{\partial x_1^{a_1}\ldots\partial x_1^{a_{n_1}}}\frac{\partial^{n_2}\grf( \vvx_2)}{\partial x_2^{b_1}\ldots\partial x_2^{b_{n_2}}}\right\rangle=\frac{\partial^{n_1+n_2}C( \vvx_1- \vvx_2)}{\partial x_1^{a_1}\ldots \partial x_1^{a_{n_1}}\partial x_2^{b_1}\ldots\partial x_2^{b_{n_2}}} \, .
\end{equation}
The indices $a$ and $b$ take values between $1$ and $d$.
The expectation values of the derivatives are given by the derivatives of the expectation value, which we take to be zero. To simplify notation, we will from now on write the derivatives of $f$ as:
\eq{
\frac{\partial^{n}\grf(\vvx)}{\partial x^{a_1}\ldots\partial x^{a_{n}}}\equiv \grf_{a_1\ldots a_n}(\vvx) \, .
}

In this paper we create the random functions by generating Taylor coefficients, $\grf_{a_1\ldots a_n}$, at a single point, $\vvx_\star$. Their joint probability distribution is just a multivariate normal distribution with a covariance matrix given by the derivatives of the covariance function at $\vvx_1=\vvx_2=\vvxs$. Of course, not all the derivatives are algebraically independent, so we only consider the coefficients $\grf_{a_1  ... a_n}$ with indices ordered such that $a_1\geq  ...\geq a_n$. This ensures that all the unique coefficients are included exactly once. 

We will consider Taylor expansions that are truncated at a finite order, $n \leq n_{\rm max}$. 
To simplify our notation, we denote the unique, ordered Taylor coefficients for any $0\leq n \leq n_{\rm max}$ indices by $f_\alpha$. 
%
%
The multivariate probability distribution function for the algebraically independent Taylor coefficients is then given by:
\begin{equation}
P(f_{\alpha})=\frac{\exp\left(-\frac12(f_{\alpha}-\mu_{\alpha})(\Sigma^{-1})_{\alpha\beta}(f_{\beta}-\mu_{\beta})\right)}{\sqrt{\det(2\pi\Sigma)}}\label{eq:pdf} \, , 
\end{equation}
where $\mu_\alpha=\langle \grf_\alpha\rangle$ is the expectation value vector and $\Sigma_{\alpha\beta}=\langle (\grf_\alpha-\mu_\alpha) (\grf_\beta-\mu_\beta)\rangle$ is the covariance matrix. Here repeated indices are a short-hand for the double-sum over $n$ and  the ordered indices for each $n$.

The key obstacle for generating realisation of the GRF from the multivariate probability distribution of equation \eqref{eq:pdf} is the need to invert the covariance matrix. When the number of independent coefficients becomes very large, numerical inversion becomes prohibitively costly. The purpose of this paper is to show how such an inversion can be circumvented for GRFs with the covariance function \eqref{eq:Gcovar} by sequential application of the marginal and conditional probability distributions for the Taylor coefficients.

 In general, if the vectors $Z_1$ and $Z_2$, are randomly distributed according to  the multivariate normal distribution,
\begin{equation}
\begin{bmatrix}
Z_1 \\
Z_2
\end{bmatrix}\sim N\left(
\begin{bmatrix}
\mu_1 \\
\mu_2
\end{bmatrix},
\begin{bmatrix}
\Sigma_{11} & \Sigma_{12}\\
\Sigma_{21} & \Sigma_{22}
\end{bmatrix}\right) \, ,
\end{equation}
where $(\mu_1,\mu_2)$ is the mean vector and $\Sigma_{ij}$ are block components of the covariance matrix, then the probability distribution for $Z_1$ obtained when marginalising over $Z_2$ is simply,
\be
Z_1 \sim N(\mu_1, \Sigma_{11}) \, .
\ee
The conditional probability distribution for the vector $Z_2$ obtained after fixing $Z_1=z_1$ is the  multivariate Gaussian distribution with expectation values and covariance given by:
\begin{align}
\mathrm E[Z_2|z_1]=&\hspace{3pt}\mu_2+\Sigma_{21}\Sigma_{11}^{-1}(z_1-\mu_1)\label{eq:EV} \, , \\[2pt] 
\mathrm{Cov}[Z_2|z_1]=&\hspace{3pt}\Sigma_{22}-\Sigma_{21}\Sigma_{11}^{-1}\Sigma_{12}\label{eq:newCov} \, .
\end{align}
We now give a simple example showing how these expressions simplify the generation of the Taylor coefficients.




\subsection{Motivational example}
\label{sec:expl}
To illustrate the main point of our method, it is convenient to initially absorb the hyperparameters $h$ and $\ell$ of equation \eqref{eq:Gcovar} into $f$ and $\vvx$. The  covariance function is then given by, 
\begin{equation}
\langle \grf(\vvx_1)\grf( \vvx_2)\rangle=e^{-(\vvx_1-\vvx_2)^2/2} \, , \label{eq:Gcovar2}
\end{equation}
and the covariances of the first few derivatives of $f$ at $\vvxs$ are:
\begin{align}
\langle \grf\grf\rangle&=1 \, , ~~~\langle f f_a \rangle= 0 \, , ~~~\langle f f_{ab} \rangle= - \delta_{ab} \, ,\\ 
\langle \grf_{a}\grf_{b}\rangle&=\delta_{ab} \, , ~~~
\langle f_a f_{bc} \rangle = 0 \, , ~~~
\langle \grf_{ab}\grf_{cd}\rangle=\delta_{ab}\delta_{cd}+\delta_{ac}\delta_{bd}+\delta_{ad}\delta_{bc} \, .
\end{align}
Note that we may obtain the covariances between derivatives of different order by changing which field derivatives act on; each time this is done, we pick up a minus sign. All covariances between even and odd derivatives vanish since the covariance function is even.
%
%

The above covariances are symmetric in all the indices, and for any given set of indices it is easy to write down the value of the covariance according to the following combinatoric rules: If any index appears an odd number of times, the covariance is zero. 
For an index appearing $n$ (even) times we get a factor of $(n-1)!!$, which is the number of unique ways they can be put together in Kroenecker deltas. The total covariance for a given set of indices is then given by the product of such factors for each index appearing in the indices. The overall sign is set by half the difference of number of indices of the two sets. For example, $\langle f_{555221} f_{995221} \rangle  = (-1)^{(6-6)/2}(4!!)^2 (2!!)^2 =256$. 

We now specialise to the  case of $d=2$. 
Since the covariances between Taylor coefficients with an even and an odd number of indices vanishes, the full covariance matrix is block diagonal. We here consider the covariance matrix for the Taylor coefficients with an even number of indices, which for the variables $f$, $f_{a_1 a_2}$, and $f_{a_1 \ldots a_4}$ is given by,
\eq{
\Sigma=
\left(
\begin{array}{c;{2pt/2pt} c c c;{2pt/2pt} c c c c c}
 1 & -1 & 0 & -1 & 3 & 0 & 1 & 0 & 3 \\ \hdashline[2pt/2pt]
 -1 & 3 & 0 & 1 & -15 & 0 & -3 & 0 & -3 \\
 0 & 0 & 1 & 0 & 0 & -3 & 0 & -3 & 0 \\ 
  -1 & 1 & 0 & 3 & -3 & 0 & -3 & 0 & -15 \\ \hdashline[2pt/2pt]

 3 & -15 & 0 & -3 & 105 & 0 & 15 & 0 & 9 \\
 0 & 0 & -3 & 0 & 0 & 15 & 0 & 9 & 0 \\
 1 & -3 & 0 & -3 & 15 & 0 & 9 & 0 & 15 \\
 0 & 0 & -3 & 0 & 0 & 9 & 0 & 15 & 0 \\
 3 & -3 & 0 & -15 & 9 & 0 & 15 & 0 & 105 
\end{array} 
\right)
\, ,
\label{eq:ExplMat}
}
where the first row/column is for $\grf$, the following three are for the $(1,1)$, $(2,1)$ and $(2,2)$ components of $f_{a_1 a_2}$, and the final five are for the $(1,1,1,1)$, $(2,1,1,1)$, etc. components of $f_{a_1 \ldots a_4}$. 

The marginal distribution for $f$ is simply given by $f \sim N(0,1)$. 
Fixing $f$, the conditional covariance matrix for $f_{a_1 a_2}$ and $f_{a_1 a_2 a_3 a_4}$ is obtained from equation \eqref{eq:newCov}:
\eq{
\Sigma=
\left(
\begin{array}{c c c;{2pt/2pt} c c c c c}
 2 & 0 & 0 & -12 & 0 & -2 & 0 & 0 \\
 0 & 1 & 0 & 0 & -3 & 0 & -3 & 0 \\
 0 & 0 & 2 & 0 & 0 & -2 & 0 & -12 \\ \hdashline[2pt/2pt]
 -12 & 0 & 0 & 96 & 0 & 12 & 0 & 0 \\
 0 & -3 & 0 & 0 & 15 & 0 & 9 & 0 \\
 -2 & 0 & -2 & 12 & 0 & 8 & 0 & 12 \\
 0 & -3 & 0 & 0 & 9 & 0 & 15 & 0 \\
 0 & 0 & -12 & 0 & 0 & 12 & 0 & 96 \\
\end{array} 
\right) \, .
}
We note that the $3\times3$ block matrix in the upper left corner is now diagonal, and the probability distribution for $f_{a_1 a_2}$ given $f$ is now given by:
\be
\left(f_{a_1 a_2}| f\right) \sim N\left(
\begin{bmatrix}
-f \\
0 \\
-f
\end{bmatrix},
\begin{bmatrix}
2 & 0 &0\\
0 & 1 & 0 \\
0 & 0 & 2
\end{bmatrix}\right) \, .
\ee 
Clearly, the second order derivatives are now independent Gaussian random numbers.

Finally, fixing $f_{a_1 a_2}$ in addition to $f$, we find that the conditional probability distribution for $f_{a_1 a_2 a_3 a_4}$ is given by:
\eq{
(f_{a_1 a_2 a_3 a_4}| f_{a_1 a_2},\, f) \sim
N\left(
\begin{bmatrix}
-3(f+2f_{11}) \\
-3f_{21} \\
-f-f_{11}-f_{22} \\
-3f_{21}\\
-3(f+2f_{22})
\end{bmatrix},
\begin{pmatrix}
 24 & 0 & 0 & 0 & 0 \\
 0 & 6 & 0 & 0 & 0 \\
 0 & 0 & 4 & 0 & 0 \\
 0 & 0 & 0 & 6 & 0 \\
 0 & 0 & 0 & 0 & 24 \\
\end{pmatrix}
\right)
 \, .
}
Since the covariance matrix  again is diagonal,  also the fourth order Taylor coefficients can be generated as independent Gaussian random numbers, without the need to diagonalise the original covariance matrix \eqref{eq:ExplMat}. 

\subsection{All orders proof}
\label{sec:proof}
In this section, we show that the method of section \ref{sec:expl} applies  for arbitrary $d$ and $n_{\rm max}$. That is, we will show that the conditional covariance matrix for the $k$:th order Taylor coefficients is diagonal, given all lower-order Taylor coefficients of the same type (even or odd). 
\subsubsection*{Additional notation}
It is convenient to introduce some additional notation that will allow us to prove our main result simultaneously for the cases of even and odd number of indices.  
The covariance matrix is again  block-diagonal, and each block further consists of the covariances of Taylor coefficients of increasingly high order, up to order $n_{\rm max}$ or $n_{\rm max}-1$. We collectively refer to the orders in the even and odd case as `levels', $i$, where $i = n/2+1$ in the even case and $i=(n+1)/2$ in the odd case. For each level, the index $\alpha_i$ runs over the ordered set of indices.
(We may replace $\alpha$ with any lower case Greek index.) The $n_i$ indices within a  set $\alpha_i$ will be labelled by the corresponding lower case Latin letters and a number, e.g.~$a_1, a_2, \ldots$.  An unordered set will be denoted $\alpha_i^u$.

For example, for even derivatives $i=1$ corresponds to $f$ (so that $n_1=0$), $i=2$ to $f_{a_1a_2}$ ($n_2=2$), etc. and for odd derivatives $i=1$ corresponds to $f_{a_1}$ ($n_1=1$), $i=2$ to $\grf_{a_1a_2a_3}$ ($n_2=3$).


It is also convenient to write the covariances as:
\begin{equation}
\langle \grf_{\alpha_i}\grf_{\beta_j}\rangle=C_{\alpha_i\beta_j} \, .
\end{equation}
Each $C_{\alpha_i\beta_j}$ will consist of a sum over Kronecker delta functions that `connect' indices in $\alpha_i$ with indices in $\beta_j$, or with other indices in $\alpha_i$. We will find it convenient to consider modified covariances,  $C_{\alpha_i\beta_j}^k$, which are obtained from $C_{\alpha_i\beta_j}$ by removing all terms with Kronecker deltas connecting either $n_1$, $n_2$, ..., or $n_k$ indices from $\alpha_i$ with the same number of indices from $\beta_j$.
We will also write,
\begin{equation}
C_{\alpha_i\beta_i}^{i-1}=D_{\alpha_i\beta_i} \, ,
\end{equation}
for which all indices in $\alpha_i$ are connected with indices in $\beta_i$, so that $D_{\alpha_i\beta_i}$ is only non-vanishing if $\alpha_i = \beta_i$, and hence, $D_{\alpha_i\beta_i}$ is a diagonal matrix. 
For example, in the even case we have:
\begin{equation}
C^1_{\alpha_2 \beta_2} = D_{\alpha_2\beta_2}=\delta_{a_1b_1}\delta_{a_2b_2}+\delta_{a_1b_2}\delta_{a_2b_1} \, .
\end{equation}
The diagonal matrix can further be written as,
\begin{equation}
D_{\alpha_i\beta_i}=\delta_{\alpha_i\beta_i}\text{Comb}(\alpha_i), \label{eq:diageq}
\end{equation}
where $\text{Comb}(\alpha_i)$ is a combinatorial factor determined by the values of the indices in $\alpha_i$, and it is the number of ways the numbers in the set can be paired up with the same numbers in an identical set. If we denote the number of times an index value $a$ appears in $\alpha_i$ by $k_a$, we then have: 
\begin{equation}
\text{Comb}(\alpha_i)=\prod_{a=1}^{d}k_a! \, .
\end{equation}
As an example, the set $\{3,3,1,1,1\}$ can be paired up with an identical ordered set in $2!\times3!=12$ ways. We also note that the total number of permutations of a set of indices $\alpha_i$ is given by:
\begin{equation}
\text{Perms}(\alpha_i)=\frac{n_i!}{\text{Comb}(\alpha_i)} \, .\label{eq:permeq}
\end{equation}
Using the same example as before, one can easily see that the set has $10$ permutations, which agrees exactly with $5!/2!3!=10$.

Finally, we will write the conditional covariance matrix for the Taylor coefficients with levels from $k$ through $n$, given values for the lower levels $1$ through $ k-1$ as
$\Sigma_{k,n}^{k-1}$.

\subsubsection*{From the first to the second level}

We now want to prove that if we specify levels $1$ through $k$, then the covariance matrix of conditional probability distributions for level $k+1$ will be diagonal. As a starting point, we write down the covariance matrix for levels $1$ through $n$ for some $n>k$:
\begin{equation}
\Sigma_{1,n}=\begin{pmatrix}
C_{\alpha_1\beta_1}  & C_{\alpha_1\beta_2}     & C_{\alpha_1\beta_3}      &   \dots     &  C_{\alpha_1\beta_{n-1}}     & C_{\alpha_1\beta_{n}}  \\
C_{\alpha_2\beta_1}  & C_{\alpha_2\beta_2}     & C_{\alpha_2\beta_3}      &   \dots     &  C_{\alpha_2\beta_{n-1}}     & C_{\alpha_2\beta_{n}}  \\
C_{\alpha_3\beta_1}  & C_{\alpha_3\beta_2}     & C_{\alpha_3\beta_3}      &   \dots     &  C_{\alpha_3\beta_{n-1}}     & C_{\alpha_3\beta_{n}}  \\
\vdots   & \vdots & \vdots &  & \vdots & \vdots  \\
C_{\alpha_{n-1}\beta_1}  & C_{\alpha_{n-1}\beta_2}     & C_{\alpha_{n-1}\beta_3}      &   \dots     &  C_{\alpha_{n-1}\beta_{n-1}}     & C_{\alpha_n\beta_{n}}\\
C_{\alpha_{n}\beta_1}  & C_{\alpha_{n}\beta_2}     & C_{\alpha_{n}\beta_3}      &   \dots     &  C_{\alpha_{n}\beta_n}     & C_{\alpha_{n}\beta_{n}}
\end{pmatrix} \, .
\end{equation} 
This can be for either the odd or the even case.
For the even case $C_{\alpha_1\beta_1}=D_{\alpha_1\beta_1} =\langle \grf\grf\rangle=1$, which can be thought of as a diagonal matrix in the set $\alpha_1$ (which only takes the value $\emptyset$). For the odd case $C_{\alpha_1\beta_1}=D_{\alpha_1\beta_1} =\langle \grf_{a_1}\grf_{b_1}\rangle=\delta_{a_1b_1}$, which again is diagonal. Thus, the lowest order Taylor coefficients can always be fixed as independent Gaussian numbers (by marginalising over all higher-order coefficients). 

Now suppose we fix $\grf_{\alpha_1}$. From equation \eqref{eq:newCov}, we then obtain the following distribution for the remaining levels:
\begin{align*}
\Sigma_{2,n}^1=&\begin{pmatrix}
C_{\alpha_2\beta_2}     & C_{\alpha_2\beta_3}      &   \dots     &  C_{\alpha_2\beta_{n-1}}     & C_{\alpha_2\beta_{n}}  \\
 C_{\alpha_3\beta_2}     & C_{\alpha_3\beta_3}      &   \dots     &  C_{\alpha_3\beta_{n-1}}     & C_{\alpha_3\beta_{n}}  \\
\vdots & \vdots &  & \vdots & \vdots  \\
 C_{\alpha_{n-1}\beta_2}     & C_{\alpha_{n-1}\beta_3}      &   \dots     &  C_{\alpha_{n-1}\beta_{n-1}}     & C_{\alpha_n\beta_{n}}\\
 C_{\alpha_{n}\beta_2}     & C_{\alpha_{n}\beta_3}      &   \dots     &  C_{\alpha_{n}\beta_n}     & C_{\alpha_{n}\beta_{n}}
\end{pmatrix}-\\[4pt]
&-\begin{pmatrix}
C_{\alpha_2\gamma_1}   \\
C_{\alpha_3\gamma_1}   \\
\vdots \\
 C_{\alpha_{n-1}\gamma_1}\\
 C_{\alpha_{n}\gamma_1}
\end{pmatrix}(D^{-1})_{\gamma_1\epsilon_1}\begin{pmatrix}
C_{\epsilon_1\beta_2}   &
C_{\epsilon_1\beta_3}   &
\dots &
 C_{\epsilon_1\beta_{n-1}}&
 C_{\epsilon_1\beta_{n}}\nt
\end{pmatrix} \, .
\end{align*}
Again, repeated indices are summed over.
In more compact notation, this can be written as:
\begin{equation}
(\Sigma_{2,n}^1)_{\alpha_i\beta_j}=C_{\alpha_i\beta_j}-C_{\alpha_{n}\gamma_1}(D^{-1})_{\gamma_1\epsilon_1}C_{\epsilon_1\beta_j} \, ,
\end{equation}
where $(\Sigma_{2,n}^1)_{\alpha_i\beta_j}$ is the conditional covariance between $\grf_{\alpha_i}$ and $\grf_{\beta_j}$, given $\grf_{\alpha_1}$.

In both the odd and even cases, $D_{\gamma_1\epsilon_1}$ is diagonal and in fact just given by $\delta_{\gamma_1\epsilon_1}$ as described above, so we find:
\begin{equation}
(\Sigma_{2,n}^1)_{\alpha_i\beta_j}=C_{\alpha_i\beta_j}-C_{\alpha_{n}\gamma_1}C_{\gamma_1\beta_j} \, .
\end{equation}
This is really just the original covariance, with all the terms mixing $n_1$ indices from $\alpha_j$ with indices from $\beta_j$ dropped. To make this obvious, we write out these terms explicitly:
\begin{equation}
(\Sigma_{2,n}^1)_{\alpha_i\beta_j}=
  \begin{cases}
    \langle \grf_{a_1...a_{2i-2}}\grf_{b_1...b_{2j-2}}\rangle-  \langle \grf_{a_1...a_{2i-2}}\grf\rangle\langle \grf\grf_{b_1...b_{2j-2}}\rangle       & \quad \text{even case} \, ,\\
     \langle \grf_{a_1...a_{2i-1}}\grf_{b_1...b_{2j-1}}\rangle-  \langle \grf_{a_1...a_{2i-1}}\grf_{c_1}\rangle\langle \grf_{c_1}\grf_{b_1...b_{2j-1}}\rangle       & \quad \text{odd case} \, .
  \end{cases}
\end{equation}
In the even case, the second term subtracts from the first all terms where the $a_i$ and $b_i$ indices do not mix, i.e.~those terms with $n_1=0$ indices from each set in mixed Kronecker deltas. In the odd case, the second term subtracts from the first all terms where only one of the $a_i$ indices and one of the $b_i$ indices are in a Kronecker delta together, i.e.~those terms with $n_1=1$ indices from each set in mixed Kronecker deltas. It then follows that we have,
\begin{equation}
(\Sigma_{2,n}^1)_{\alpha_2\beta_2}=C_{\alpha_2\beta_2}^1=D_{\alpha_2\beta_2}=\delta_{\alpha_2\beta_2}\text{Comb}(\alpha_2) \, . \label{eq:Sigma12}
\end{equation}
Equation \eqref{eq:Sigma12} implies that the second order coefficients of the conditional probability distribution are statistically independent, and can be generated without explicitly inverting a non-trivial covariance matrix (when marginalising over higher levels).

To further  illustrate  equation  \eqref{eq:Sigma12}, we see that in the even case it  explicitly corresponds to:
\begin{equation}
(\Sigma_{2,n}^1)_{\alpha_2\beta_2}=\langle \grf_{a_1a_2}\grf_{b_1b_2}\rangle-  \langle \grf_{a_1a_2}\grf\rangle\langle \grf\grf_{b_1b_2}\rangle  =\delta_{a_1b_1}\delta_{a_2b_2} +\delta_{a_1b_2}\delta_{a_2b_1} \, ,
\end{equation}
which follows from $ \langle \grf_{a_1a_2}\grf\rangle=-\delta_{a_1a_2}$. In the odd case, we have,
\begin{align*}
(\Sigma_{2,n}^1)_{\alpha_2\beta_2}&=\langle \grf_{a_1a_2a_3}\grf_{b_1b_2b_3}\rangle-  \langle \grf_{a_1a_2a_3}\grf_{c_1}\rangle\langle \grf_{c_1}\grf_{b_1b_2b_3}\rangle \\
&=\delta_{a_1b_1}\delta_{a_2b_2}\delta_{a_3b_3} +\delta_{a_1b_1}\delta_{a_2b_3}\delta_{a_3b_2}+\delta_{a_1b_2}\delta_{a_2b_1}\delta_{a_3b_3}+ \delta_{a_1b_2}\delta_{a_3b_2}\delta_{a_3b_1}\\&\quad+ \delta_{a_1b_3}\delta_{a_2b_2}\delta_{a_3b_1} +\delta_{a_1b_3}\delta_{a_3b_2}\delta_{a_2b_3}\nt 
\label{eq:Sigma12odd}
\, .
\end{align*}
Here we have used,
\begin{equation}
\langle \grf_{a_1a_2a_3}\grf_{c_1}\rangle=-\delta_{a_1a_2}\delta_{a_3c_1}-\delta_{a_1a_3}\delta_{a_2c_1}-\delta_{a_2a_3}\delta_{a_1c_1} \, ,
\end{equation}
and a similar expression for $\langle \grf_{c_1}\grf_{b_1b_2b_3}\rangle$. 
The second term of the top line of equation \eqref{eq:Sigma12odd} will cancel out the terms where only one delta contains an index each from both sets, and since every term must have either one or three such deltas (there are three indices in each set), it follows that the only terms that remain are those in which all the indices are mixed.

\subsubsection*{For pedagogical reasons: from the second to the third  level}

To illustrate the structure of our general recursive proof, we here consider the less trivial step of the conditional covariance after fixing both $\grf_{\alpha_1}$ and 
 $\grf_{\alpha_2}$. The covariance matrix for the higher levels is then given by,
\begin{equation}
(\Sigma_{3,n}^2)_{\alpha_i\beta_j}=C^1_{\alpha_i\beta_j}-C^1_{\alpha_{i}\gamma_2}(D^{-1})_{\gamma_2\epsilon_2}C^1_{\epsilon_2\beta_j},
\end{equation}
with $3\leq i,j\leq n$. Now, using the expression for the diagonal matrix given in equation \eqref{eq:diageq}, we can write this as,
\begin{equation}
(\Sigma_{3,n}^2)_{\alpha_i\beta_j}=C^1_{\alpha_i\beta_j}-\sum_{\gamma_2}C^1_{\alpha_{i}\gamma_2}C^1_{\gamma_2\beta_j}/\text{Comb}(\gamma_2) \, ,
\end{equation}
where the sum is over ordered sets of indices. 

To  generalise the  argument of the previous section, we first rewrite this in terms a sum over unordered indices. This will overcount the index sets by a factor of how many permutations there are of them, so in every term we need to divide by the number of permutations:
\begin{align*}
(\Sigma_{3,n}^2)_{\alpha_i\beta_j}&=C^1_{\alpha_i\beta_j}-\sum_{\gamma^u_2}C^1_{\alpha_{i}\gamma^u_2}C^1_{\gamma^u_2\beta_j}/(\text{Comb}(\gamma_2)\text{Perms}(\gamma_2))\\
&=C^1_{\alpha_i\beta_j}-\sum_{\gamma^u_2}C^1_{\alpha_{i}\gamma^u_2}C^1_{\gamma^u_2\beta_j}/n_2! \, , \nt \label{eq:2to3}
\end{align*}
where in the last step we used equation \eqref{eq:permeq}. We recall that $C^1_{\alpha_{i}\gamma^u_2}$ includes no  terms with $n_1$ indices connecting $\alpha_i$ and $\gamma_2^u$. Since  these cannot be connected by fewer than $n_1$ indices (recall that $n_1^\text{even}=0$ and $n_1^\text{odd}=1$), they must be connected by  more than $n_1$ indices.
More generally, $C_{\alpha_i\beta_j}^k$ only contains terms in which 
 $n\in\{n_{k+1},n_{k+2},...\}$  indices from two sets $\alpha_i$ and $\beta_j$ are paired up in Kronecker deltas.
 This is because there must be even numbers of indices left in the sets $\alpha_i$ and $\beta_j$, and all the $n_l$ differ by multiples of two. E.g.~for the odd case $\alpha_3$ has $n_3=5$ indices and only $n_1=1$, $n_2=3$ or $n_3=5$ indices from $\alpha_3$ can be paired up with indices from another set in $C_{\alpha_5\beta_j}$. This then tells us that in $C^1_{\alpha_{i}\gamma^u_2}$, all $n_2$ of the indices from the set $\gamma_2^u$ are together in deltas with some indices from $\alpha_i$.

 Note also that the terms in the $C^{1}_{\alpha_i\beta_j}$ may be negative, but the relative sign between the two terms in equation \eqref{eq:2to3} is always the same: the sign in front of the Kronecker deltas in $C^1_{\alpha_i\beta_j}$ is $(-1)^{i-j}$ and in $C^1_{\alpha_{i}\gamma_2}C^1_{\gamma_2\beta_j}$ it is $(-1)^{i-1}(-1)^{1-j}=(-1)^{i-j}$. This obviously also holds if we replace $1$ with any other index $k$.


Now consider a subset of $\alpha_i$ containing $n_2$ indices. In $C^1_{\alpha_i,\gamma_2^u}$ the indices in this subset will be paired together with the indices in $\gamma_2^u$ into deltas in $n_2!$ different ways. The same applies for any given subset of $\beta_j$ containing $n_2$ indices in $C^1_{\gamma_2^u,\beta_j}$. When these two terms are multiplied together, each combination of the $\alpha_i$ subset and $\gamma_2^u$ indices will multiply the $\beta_j$ subset in all combinations with the $\gamma_2^u$ indices, giving all $n_2!$ combinations of the $\alpha_i$ and $\beta_j$ subsets paired into deltas. Every combination will therefore appear $n_2!$ times when all the terms are added together, cancelling out the factor of $1/n_2!$ in equation \eqref{eq:2to3}. The second term of equation \eqref{eq:2to3} will then subtract off all the terms with $n_2$ indices from each set $\alpha_i$ and $\beta_j$ mixed in deltas. What remains  is then the initial covariance matrix minus the terms connecting $n_1$ or $n_2$ indices from the different sets. We thus have,
\begin{equation}
(\Sigma_{3,n}^2)_{\alpha_i\beta_j}=C^2_{\alpha_i\beta_j} \, , \label{eq:Sigma23}
\end{equation}
from which it  immediately follows that:
\begin{equation}
(\Sigma_{3,n}^2)_{\alpha_3\beta_3}=C^2_{\alpha_3\beta_3}=D_{\alpha_3\beta_3} \, .
\end{equation}
The third level coefficients therefore become statistically independent random variables if all the lower level coefficients are known.

\subsubsection*{Proof by induction}
We may now recursively  show that  the above procedure and the diagonalisation of the conditional covariance matrix  hold to any level. 
We first assume that,
\begin{equation}
(\Sigma_{k,n}^{k-1})_{\alpha_i\beta_j}=C^{k-1}_{\alpha_i\beta_j} \, , \label{eq:assump}
\end{equation}
where, again,  the superscript $k-1$ on the $C^{k-1}_{\alpha_i\beta_j}$ means that terms connecting $n_1$, $n_2$, ... or $n_k$ indices between  $\alpha_i$ and $\beta_j$  are not present in the covariances. Equation \eqref{eq:assump} holds for $k=2$ (cf.~equation \eqref{eq:Sigma12}) and $k=3$ (cf.~equation \eqref{eq:Sigma23}). We would now like to show that equation \eqref{eq:assump} holds for $k \to k+1$.

Upon fixing $\grf_{\alpha_k\beta_k}$ (as well as the lower levels), the conditional covariance matrix for the higher levels is given by,
\begin{align*}
(\Sigma_{k+1,n}^{k})_{\alpha_i\beta_j}&=C^{k-1}_{\alpha_i\beta_j}-C^{k-1}_{\alpha_{i}\gamma_k}(D^{-1})_{\gamma_k\epsilon_k}C^{k-1}_{\epsilon_k\beta_j}\\
&=C^{k-1}_{\alpha_i\beta_j}-\sum_{\gamma_k}C^{k-1}_{\alpha_{i}\gamma_k}C^{k-1}_{\gamma_k\beta_j}/\text{Comb}(\gamma_k)\\
&=C^{k-1}_{\alpha_i\beta_j}-\sum_{\gamma_k^u}C^{k-1}_{\alpha_{i}\gamma^u_k}C^{k-1}_{\gamma^u_k\beta_j}/n_k!\nt \label{eq:ktokp1},
\end{align*}
where we have taken precisely the same steps that led us to equation \eqref{eq:2to3}. 

The superscript $k-1$ on the $C^{k-1}_{\alpha_{i}\gamma^u_k}$ indicates  that 
it contains no terms with  $n_1$, $n_2$, ... $n_{k-1}$ deltas with one index each from $\alpha_i$ and one from $\gamma_k^u$. No term can have fewer than $n_1$ indices 
connecting $\alpha_i$ and  $\gamma_k^u$,
 and since there are $n_k$ indices in $\gamma_k^u$ it then follows that they all must be connected with an index in $\alpha_i$.


We now consider a subset of $\alpha_i$ containing $n_k$ indices. In $C^{k-1}_{\alpha_i,\gamma_k^u}$ the indices in this subset will be paired together with the indices in $\gamma_k^u$ into deltas in $n_k!$ different ways. The same applies for any given subset of $\beta_j$ containing $n_k$ indices in $C^{k-1}_{\gamma_k^u,\beta_j}$. When these two terms are multiplied together, each combination of the $\alpha_i$ subset and $\gamma_k^u$ indices will multiply the $\beta_j$ subset in all combinations with the $\gamma_k^u$ indices, giving all $n_k!$ combinations of the $\alpha_i$ and $\beta_j$ subsets paired into deltas. Every combination will therefore appear $n_k!$ times when all the terms are added together, cancelling out the factor of $1/n_k!$ in equation \eqref{eq:ktokp1}. The second term in equation \eqref{eq:ktokp1} will then subtract off all the terms with $n_k$ indices from each set $\alpha_i$ and $\beta_j$ mixed in deltas. What remains is then the initial covariance matrix connecting at least $n_{k+1}$ indices from the sets $\alpha_i$ and $\beta_j$. That is, we have:
\begin{equation}
(\Sigma_{k+1,n}^{k})_{\alpha_i\beta_j}=C^{k}_{\alpha_i\beta_j} \, .
\end{equation}
We conclude that this expression holds for any level. In particular, this implies that,
\be
(\Sigma_{k+1,n}^{k})_{\alpha_{k+1} \beta_{k+1}}=C^{k}_{\alpha_{k+1}\beta_{k+1}} = D_{\alpha_{k+1}\beta_{k+1}} = 
\delta_{\alpha_{k+1} \beta_{k+1}}\, {\rm Comb}(\alpha_{k+1}) \, ,
\label{eq:final}
\ee
from which it follows that the $(k+1)$:st level Taylor coefficients can be generated as statistically independent Gaussian variables if all the the lower level coefficients are known.

\subsubsection*{The expectation values}

The only thing that remains to do now is calculate how the expectation values shift as we fix  the Taylor coefficients level by level. This just involves evaluating equation \eqref{eq:EV} in the general case. If we fix $\grf_{\alpha_k}$, then the expectation values for the higher levels ($i>k$) are changed by,
\begin{equation}
\mu_{\alpha_i}\to\mu_{\alpha_i}+C^{k-1}_{\alpha_i\beta_k}(D^{-1})_{\beta_k\gamma_k}(\grf_{\gamma_k}-\mu_{\gamma_k}) \, ,
\end{equation}
where $\mu_{\gamma_k}$ will have been determined by earlier measurements. 
The elements of  the `shift matrix' matrix,
\begin{equation}
E_{\alpha_i\beta_k}=C^{k-1}_{\alpha_i\gamma_k}(D^{-1})_{\gamma_k\beta_k}=C^{k-1}_{\alpha_i\beta_k}\text{Comb}(\beta_k)^{-1} \, ,\label{eq:Edef}
\end{equation}
can be deduced with combinatorics.

We start by considering $C^{k-1}_{\alpha_i\beta_k}$. Again, every index in $\beta_k$ must be connected with an index in $\alpha_i$, and as before, we are dealing with ordered sets of indices.
  For a given $\alpha_i$ and $\beta_k$, $C^{k-1}_{\alpha_i\beta_k}$ will be given by the product of two numbers: the number of ways we can pair up indices in $\alpha_i$ with those $\beta_k$, and the number of ways the remaning indices can be paired up with each other, with an overall sign given by $(-1)^{i-k}$. To get $E_{\alpha_i\beta_k}$ one then just divides by $\text{Comb}(\beta_k)$. 

\subsection{Summary}

In sum, we have shown that realisations of the Taylor coefficients of the Gaussian random field with the covariance function \eqref{eq:Gcovar2} can be generated sequentially, in an order-by-order fashion, as independent Gaussian random numbers with the diagonal covariance matrix: 
\be
(\Sigma_{k+1,n}^{k})_{\alpha_{k+1} \beta_{k+1}}=C^{k}_{\alpha_{k+1}\beta_{k+1}} = D_{\alpha_{k+1}\beta_{k+1}}
\ee
This obviates the need to numerically invert large covariance matrices. 
%
%

The effect of the non-vanishing covariances the different orders is here encoded in the mean values, which  shift at each step by,
\eq{
\mu_{\alpha_i}\to\mu_{\alpha_i}+E_{\alpha_i\beta_k}(\grf_{\beta_k}-\mu_{\beta_k}) \, ,
}
for $i>k$, where $k$ corresponds to the order of the generated Taylor coefficients, and with $E_{\alpha_i\beta_k}$ defined in equation \eqref{eq:Edef}.
The elements of the matrices $D_{\alpha_i\beta_i}$ and $E_{\alpha_i\beta_k}$ are simple combinatorial factors that depend on the set of indices. 



\section{Applications}
\label{sec:apps}

In this section, we briefly present two applications of our construction. 
An accompanying  Mathematica notebook with examples is available at \href{https://doi.org/10.17863/CAM.22859}{https://doi.org/10.17863/CAM.22859}.


\subsection{Random, high-dimensional potential energy landscapes}

The first  application of our method is the efficient generation of random functions locally around a point $\vvxs \in \mathbb{R}^d$. The statistical independence of the Taylor coefficients when generated order-by-order allows for the study of $d\gg 1$ or $n_{\rm max} \gg 1$. This way our method can  be competitive with other ways of generating explicit GRFs (e.g.~through the generation of independent Fourier coefficients)  when the explicit function is only needed in a moderately small neighbourhood around $\vvxs$, and the covariance function is Gaussian.  
Moreover, a key benefit of our method is that it permits the generation of the function around special points: for example, by fixing the linear-order Taylor coefficients  to $f_a=0$, one can easily generate the otherwise random function  in neighbourhoods around critical points. Moreover, one may fix the  Taylor coefficients up to second order, i.e.~$f$, $f_a$, $f_{ab}$, to permit the generation of random potentials around minima, maxima, and saddle-points. We here briefly review how this has allowed us to explicitly address the problem of `manyfield' cosmological inflation  in GRF scalar potentials for the first time \cite{Bjorkmo:2017nzd}.

The explicitly generated GRFs can be interpreted as physical potential energy densities, $V$, that are functions of the $d$ scalar fields $\phi \in \mathbb{R}^d$:\footnote{The scalar fields are real-valued, dynamical  functions over the four-dimensional spacetime.}  $V = V(\phi_1,\ldots \phi_d)$. The hyper-parameters of the covariance function \eqref{eq:Gcovar} are dimensionful and may be written $\ell = \Lambda_{\rm h}$ and $h = \Lambda_{\rm v}^4$ so that:
\eq{
C(\phi_1,\phi_2)=\Lv^8e^{-(\phi_1-\phi_2)^2/2\Lh^2} \, .
}
The energy scale $\Lh$ sets the coherence length of the scalar potential, and is physically interpreted as the high-energy cut-off below which the field theory is expected to be valid. The energy scale $\Lv$ sets the `vertical scale' of the potential energy. 
The scalar potential generated to  order $n_\text{max}$ at the point $\phi_a=0$ is given by:
\eq{
V(\phi)=\Lv^4\left[\tilde V_0+\sum_{n=1}^{n_\text{max}}\frac1{n!}\tilde V_{a_1...a_n}\phi_{a_1}...\phi_{a_n}/\Lh^{n}\right] \, ,
}
where the $\tilde V_{\alpha}$ are the dimensionless Taylor coefficients. 

Cosmic inflation is a hypothetical period of accelerated expansion in the early universe that provides the leading theory for the primordial origin of cosmic structure. 
Current observations are consistent with inflation being driven by a single field, but there are good theoretical and phenomenological reasons to consider a more general situation with multiple dynamically important fields in the early universe (for two recent reviews, see \cite{Byrnes:2010em, Baumann:2014nda}). 

During slow-roll inflation, the  scalar fields  evolve over an unusually flat region of the potential energy, and  inflation can  be sustained for a sufficiently long period 
if the local `slow-roll parameters' satisfy,
\be
\epsilon_V = \frac{\Mpl^2}{2} \frac{V_a V_a}{V^2} \ll 1 \, ,
~~~|\eta_V| = \left|\Mpl^2 \frac{{\rm min(Eig(}V_{ab}))}{V} \right| \ll 1 \, ,
\label{eq:SR}
\ee
where $\Mpl$ denotes the reduced Planck mass. These conditions are rarely hold for random points in the potential energy landscapes constructed from GRFs, 
and satisfying them proved to be a major obstacle for early attempts of using Fourier representation of GRFs to study multi-field inflation \cite{Frazer:2011tg, Frazer:2011br}. 

By contrast, our method is well-suited to address this problem. By fixing  $\tilde V_0$, $\tilde V_a$ and $\tilde V_{ab}$ by hand as discussed in \cite{Bjorkmo:2017nzd}, the conditions \eqref{eq:SR} can be  locally satisfied  while the potential interactions at higher orders are non-trivial.  During inflation, the fields only explore a small region around $\phi =0$ (i.e.~the field excursion satisfies $|\Delta \phi| \ll \Lh$), and it suffices to expand the potential to $n_{\rm max} =5$. This potential is still complex enough to capture non-trivial multifield dynamics, and is appropriate for accurately computing cosmological observables generated during inflation. 
For example, for $d=100$, we fix the 5151 terms $\tilde V_0$, $\tilde V_a$ and $\tilde V_{ab}$ by hand,\footnote{In practice, we work in the eigenbasis of $\tilde V_{ab}$, so we only need to fix $2d+1$ coefficients.} and generate  the remaining 
 96,555,395
interaction terms sequentially as  independent Gaussian numbers. 

The resulting cosmology from these models is striking: despite their complexity, their predictions are remarkably simple and robust. These models are consistent with current observations, but may be tested observationally and ruled out by future cosmic microwave background experiments.  Moreover, an observable sometimes regarded as the key test of multifield inflation (the amplitude of `local shape' non-Gaussianities in the three-point correlation function of the primordial perturbations) is naturally very small, and  stringent observational limits from future experiments will not substantially constrain these general models of multifield inflation. Finally, the predictions from these models  are in excellent agreement with manyfield models constructed through a very different method, using non-equilibrium random matrix theory, in \cite{Marsh:2013qca, Dias:2016slx, Dias:2017gva,  Wang:2016kzp, Pedro:2016jyd}. This can be understood to be a consequence of eigenvalue repulsion of the Hessian matrix, which drive the predictions and is  common  to both constructions, and indeed, much broader classes of random potentials. For more details, see \cite{Bjorkmo:2017nzd, Dias:2017gva}.

Our method can also be used to study potentials over distances of a few $\Lh$ by taking $n_{\rm max} \gg 1$ (as illustrated in Figure \ref{fig:2dpotential}). This 
provides novel opportunities to address a rich class of problems involving multifield dynamics in  random potential energy landscapes. 

\subsection{Model selection for Gaussian process regression}
The results derived in this paper can also be useful when training Gaussian random processes with square exponential covariance functions. 
More specifically, it can be applied to simplify the regression problem mentioned in the introduction: given training data in the form of Taylor coefficients of $f$ at $\vvxs$ to order $n_{\rm max}$, we may determine the hyperparameters $h$, $\ell$  and $\bar f$ by maximising the log likelihood. In this section, we show in detail how the log likelihood can be written as a sum over levels, thereby making it possible to constrain the hyperparameters without inverting the covariance matrix.  


The log marginal likelihood  is given by:
\eq{
\ln P(f_\alpha|h,\ell, \bar f)=-\frac12f_\alpha \Sigma^{-1}_{\alpha\beta} f_\beta-\frac12\ln|2\pi\Sigma| \, .
}
In general, this is a very complicated function of $h$, $\ell$ and $\bar f$, but we can simplify it substantially by using the algebraic properties proven in this paper. 
First, we can use 
the definition of conditional probabilities to write:
\be
P(f_{\alpha})=\left[ \left(\prod_{i=2}^{n} P(f_{\alpha_i}|f_{\alpha_{i-1}},..)\right)P(f_{\alpha_1}) \right]_{\rm even}
\times
\left[ \left(\prod_{i=2}^{n} P(f_{\alpha_i}|f_{\alpha_{i-1}},..)\right)P(f_{\alpha_1}) \right]_{\rm odd}
\, , 
\ee
where we for simplicity of notation have suppressed the dependence on the hyperparameters.
From equation \eqref{eq:final}, the conditional probability distributions can be further simplified to,
\eq{
P(f_{\alpha_i}|f_{\alpha_{i-1}},..)=\frac1{\sqrt{|2\pi h^2 \ell^{-2n_i}D_{\alpha_i\beta_i}|}}\exp\left(-\frac12h^{-2}\ell^{2n_i}(f_{\alpha_i}-\mu_{\alpha_i})D_{\alpha_i\beta_i}^{-1}(f_{\beta_i}-\mu_{\beta_i})\right) \, ,
\label{eq:Pcond}
}
where $\mu_{\alpha_i}$ is a function of $f_{\alpha_j}$ for $j<i$ as discussed towards the end of \S\ref{sec:proof}. Denoting the number of independent derivatives at level $i$ by $d_i$, we  find,
\begin{align*}
\ln P(f_{\alpha_i}|f_{\alpha_{i-1}},..)&=-\frac12h^{-2}\ell^{2n_i}(f_{\alpha_i}-\mu_{\alpha_i})D_{\alpha_i\beta_i}^{-1}(f_{\beta_i}-\mu_{\beta_i})\\
&\hspace{12pt}-d_i(\ln h-n_i\ln \ell)-\frac{d_i}{2}\ln(2\pi)-\frac12\Tr\ln(D_{\alpha_i\beta_i}) \, .\nt
\end{align*}
The log marginal likelihood is then given by the simple sum,
\begin{align*}
\ln P(f_\alpha|h,\ell,\bar f)=\sum^{i_{\rm max}}_{\stackrel{i=1}{\rm even,\, odd}}\bigg[&-\frac12h^{-2}\ell^{2n_i}(f_{\alpha_i}-\mu_{\alpha_i})D_{\alpha_i\beta_i}^{-1}(f_{\beta_i}-\mu_{\beta_i})\\
&-d_i(\ln h-n_i\ln \ell)-\frac{d_i}{2}\ln(2\pi)-\frac12\Tr\ln(D_{\alpha_i\beta_i})\bigg],\nt
\label{eq:probsum}
\end{align*}
which runs over both even and odd derivatives.
The remaining  complication of \eqref{eq:probsum} is the mean values $\mu_{\alpha_i}$, which are determined iteratively as: 
\eq{
\mu_{\alpha_{i}}=\sum_{j=1}^{i-1}\ell^{n_j-n_i}E_{\alpha_{i}\alpha_j}(f_{\alpha_j}-\mu_{\alpha_j}) \, ,
}
with $\mu_{\alpha_1}=\bar f$ in the even case and $\mu_{\alpha_1}=0$ in the odd case. The shift matrix $E_{\alpha_i \alpha_j}$ was defined in equation \eqref{eq:Edef}. Using this expression, the expectation values $\mu_{\alpha_i}$ can be determined without inverting any matrix, which makes 
numerical evaluations fast. Derivatives of the mean values are  given recursively by,
\begin{align}
\frac{\partial\mu_{\alpha_{i}}}{\partial \ell}&=\sum_{j=1}^{i-1}\bigg[(n_j-n_i)\ell^{n_j-n_i-1}E_{\alpha_{i}\alpha_j}(f_{\alpha_j}-\mu_{\alpha_j})-\ell^{n_j-n_i}E_{\alpha_{i}\alpha_j}\frac{\partial\mu_{\alpha_j}}{\partial \ell}\bigg] \, ,\\
\frac{\partial\mu_{\alpha_{i}}}{\partial \bar f}&=\sum_{j=1}^{i-1}-\ell^{n_j-n_i}E_{\alpha_{i}\alpha_j}\frac{\partial\mu_{\alpha_j}}{\partial \bar f} \, .
\end{align}
The mean values $\mu_{\alpha_i}$ do not depend on $h$. The only non-zero starting value at first order is $\partial \mu_{\alpha_1}/\partial \bar f=1$ in the even case.

To find the Bayesian best-fit parameters, we optimise the log marginal likelihood with respect to all the hyperparameters:
\begin{align*}
\frac\partial{\partial \ell}\ln P(f_\alpha|h,\ell, \bar f)&=\sum_i\bigg[-n_ih^{-2}\ell^{2n_i-1}(f_{\alpha_i}-\mu_{\alpha_i})D_{\alpha_i\beta_i}^{-1}(f_{\beta_i}-\mu_{\beta_i})
\\
&\hspace{1.25cm}+h^{-2}\ell^{2n_i}(f_{\alpha_i}-\mu_{\alpha_i})D_{\alpha_i\beta_i}^{-1}\frac{\partial\mu_{\beta_i}}{\partial \ell}+d_in_i \ell^{-1}\bigg]\nt
\label{eq:dcond1}
\\
\frac\partial{\partial h}\ln P(f_\alpha|h,\ell,\bar f)&=\sum_i\bigg[h^{-3}\ell^{2n_i}(f_{\alpha_i}-\mu_{\alpha_i})D_{\alpha_i\beta_i}^{-1}(f_{\beta_i}-\mu_{\beta_i})-d_ih^{-1}\bigg]\nt
\label{eq:dcond2}
\\
\frac\partial{\partial \bar f}\ln P(f_\alpha|h,\ell,\bar f)&=\sum_ih^{-2}\ell^{2n_i}(f_{\alpha_i}-\mu_{\alpha_i})D_{\alpha_i\beta_i}^{-1}\frac{\partial\mu_{\beta_i}}{\partial \bar f}\nt. \label{eq:dcond3}
\end{align*}
Evidently, the only matrices that appear in this problem are  the diagonal $D_{\alpha_i\beta_i}$ and the  shift matrix $E_{\alpha_i\beta_j}$, but these are easy to compute from combinatorics, cf.~equations \eqref{eq:diageq} and \eqref{eq:Edef}. Moreover, once computed for a given $d$ and $n_{\rm max}$, they can be re-used for any training data set. 
Clearly, this model selection problem is controlled by the mean values $\mu_{\alpha_i}$ and their derivatives, and require no inversion of large matrices.

\begin{figure}
    \centering
    \begin{subfigure}{0.48\textwidth}
    \centering
    \includegraphics[width=1\textwidth]{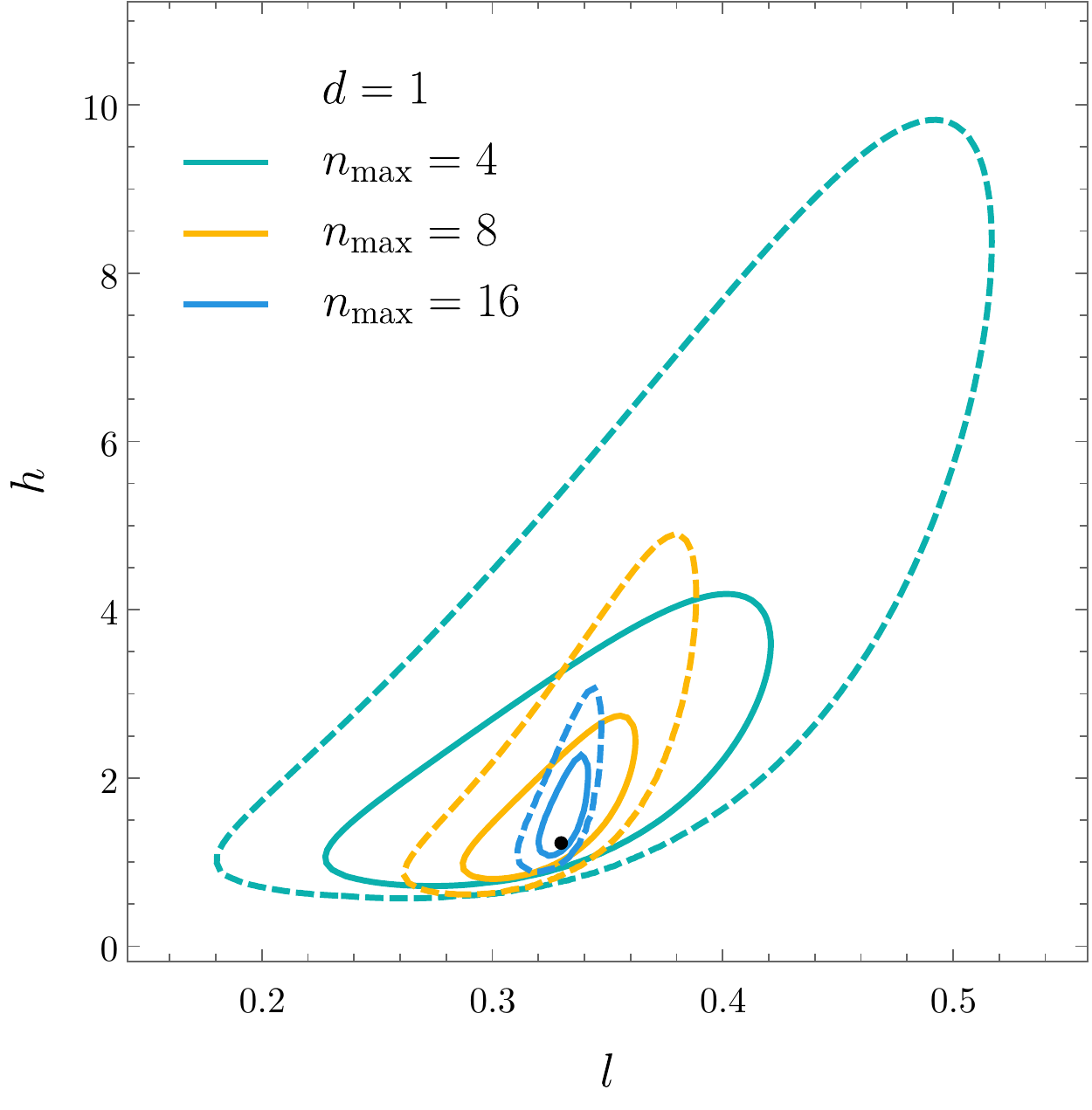}
    \end{subfigure}
    ~ 
    \begin{subfigure}{0.48\textwidth}
         \includegraphics[width=1\textwidth]{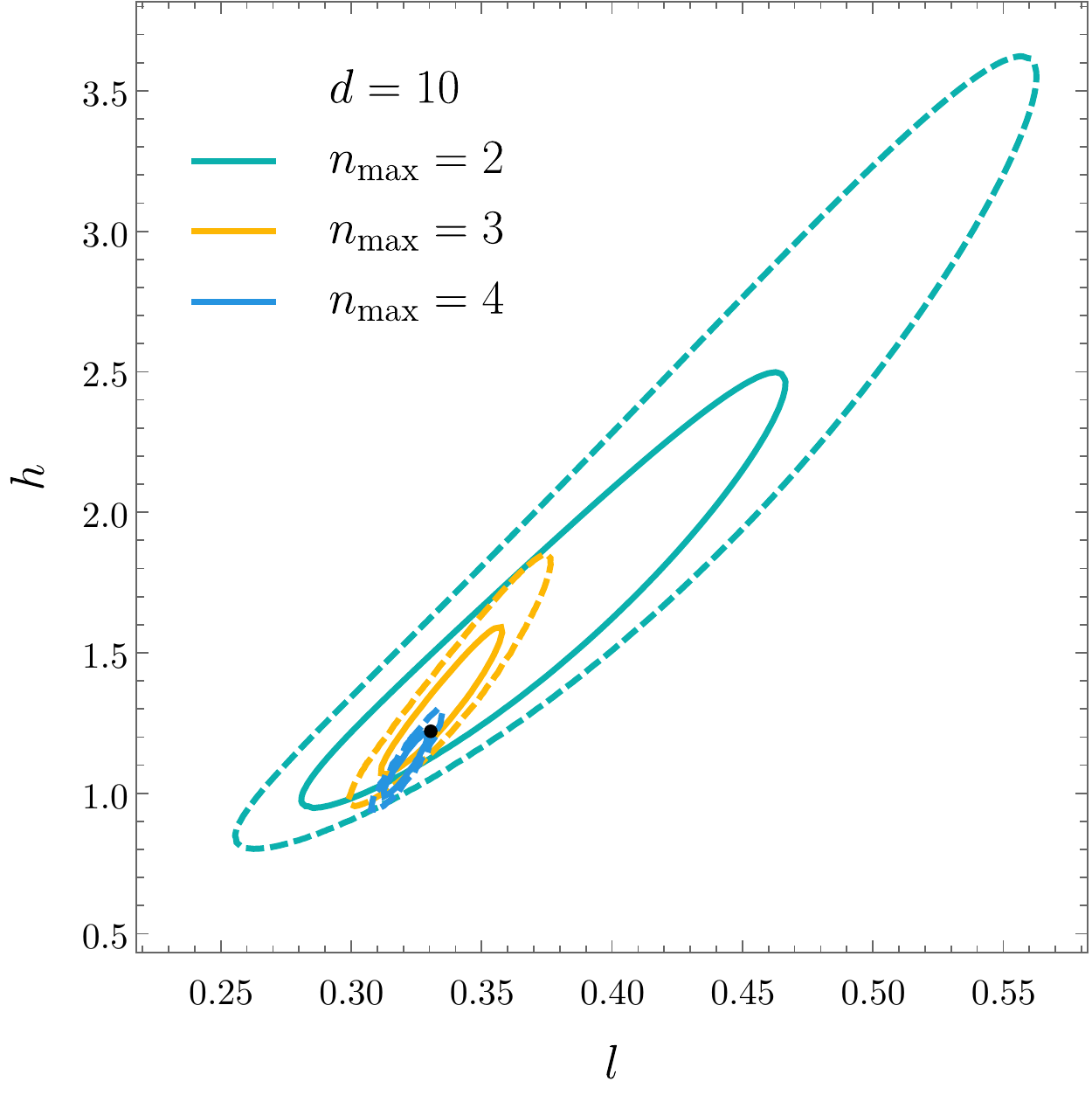}

    \end{subfigure}
    \caption{Likelihood contours for the hyperparameters at the 68\% and 95\% confidence levels obtained from random realisations of the GRF. The true hyperparameter values are $(h, \ell) = (1.22, 0.33)$ as indicated by the black dot, and we have fixed $\bar f=0$.}
    \label{fig:d1d10examples}
\end{figure}

We illustrate the application of this method for randomly generated data in Figure \ref{fig:d1d10examples}. The hyperparameters $h$ and $\ell$ can  be determined rather accurately given Taylor coefficients to sufficiently high order. Heuristically, the oblongated shape of the confidence contours can  be understood  to follow from the appearance of the pre-factors $h^{-2}\ell^{2n_i}$ in the exponent of equation \eqref{eq:Pcond}. While the degeneracy $h\to \lambda h$, $\ell \to \lambda^{1/n_i} \ell$ certainly is broken in several ways (by the $\ell$ dependence of $\mu_{\alpha_i}$, the differentiation in equations \eqref{eq:dcond1}--\eqref{eq:dcond3}, and by factors with different $n_i$),  the hyperparameters are most strongly constrained  when $\ell$ and $h$ are not both increased or decreased from the best-fit value.


\section{Conclusions}
\label{sec:concl}
In this paper, we have presented a new method for simplifying the PDF of the Taylor coefficients of GRFs with a Gaussian covariance function. By sequentially applying the marginal and conditional probability distributions, we have shown that the covariance matrices for the Taylor coefficients at every step become diagonal. This result holds  for any dimension of the GRF and to any order in the Taylor coefficients. This simplification essentially trivialises the evaluation of the probability distribution of the Taylor coefficients, which depends on the inverse of the covariance matrix. 

We have shown that this method can have several interesting applications. GRFs constructed this way can be used as models of complicated potential energy functions, and we have shown how this can be used to explicitly study  cosmic inflation in theories many interacting fields. Moreover, GRFs with Gaussian covariance functions appear very frequently in machine learning applications. We have demonstrated that our method can be used to simplify the regression problem of determining the hyperparameters, given a training data set that consists of the local Taylor coefficients of the GRF.

Accompanying this paper, we provide a Mathematica notebook containing the explicit examples of these applications.

Our method has several limitations. The algebraic simplifications that we have identified are only applicable to isotropic Gaussian covariance functions, which constitute a special, albeit commonly considered, class of GRFs. We know of no generalisations to more general classes of covariance functions, including those constructed as sums of independent Gaussian functions.   Moreover, since our method is based on the local Taylor coefficients, it becomes cumbersome to use for describing the potential over  field displacements of many $\ell$. Finally, training data sets in machine learning applications are often noisy, which can make it challenging to determine the Taylor coefficients to a sufficiently high order. 

Nevertheless, we expect our method to offer useful simplifications to a wide variety of  practical applications of Gaussian random fields and Gaussian processes.

\section*{Acknowledgements}
T.B.~is funded by an STFC studentship at DAMTP, University of Cambridge.
D.M.~is supported by a Stephen Hawking Advanced Fellowship at the Centre for Theoretical Cosmology, DAMTP, University of Cambridge.



%


\bibliographystyle{JHEP}
\bibliography{GRFrefs}


%
%



\end{document}